\documentclass[prd,twocolumn,showpacs,showkeys,preprintnumbers,amsmath,amssymb,nofootinbib]{revtex4}
\usepackage{amsmath}
\usepackage{amssymb}
\usepackage{amsfonts}
\usepackage{eucal}
\newcommand{\be}{\begin{equation}}
\newcommand{\bse}{\begin{subequations}}
\newcommand{\ese}{\end{subequations}}
\newcommand{\bea}{\begin{eqnarray}}
\newcommand{\eea}{\end{eqnarray}}
\newcommand{\ba}{\begin{array}}
\newcommand{\ea}{\end{array}}
\newcommand{\ee}{\end{equation}}

\def\nc{noncommutative\ }
\def\ncy{noncommutativity\ }

\def\fillbox#1{\hbox to #1{\vbox to #1{\vfil}\hfil}}

\makeatletter
\@addtoreset{equation}{section}

\makeatother
\begin{document}

\preprint{
%           hep-ph/0511323 \cr
           IPM/P-2005/082\cr
           HIP-2005-51/TH\cr
}
%{\vskip .2cm}
\vspace*{1mm}

\title{ Noncommutativity of Space and Rotation of Polarization of Light\\ in a Background Magnetic Field
}

\author{\bf M. Chaichian$^{1}$, M.M. Sheikh-Jabbari$^{2}$, A. Tureanu$^{1}$}
\affiliation{$^1$Department of Physical Sciences, High Energy
Physics Division, University of Helsinki\\and Helsinki Institute of
Physics, P.O.Box 64, FIN-00014 Helsinki, Finland\\
{\tt{masud.chaichian, anca.tureanu@helsinki.fi}}}
%\email{masud.chaichian, anca.tureanu@helsinki.fi}
\affiliation{$^2$ Institute for Studies in Theoretical Physics and
Mathematics (IPM), P.O.Box 19395-5531, Tehran, Iran\\
{\tt{jabbari@theory.ipm.ac.ir}}}

\begin{abstract}
Recently the PVLAS collaboration has reported the observation of
rotation of polarization of light propagating in a background
magnetic field. In this letter we explore the possibility that such
a rotation is a result of noncommutativity in the background
space-time. To explain the reported polarization rotation within
noncommutative QED we need the noncommutativity parameter
$\theta\simeq (30\ {\rm GeV})^{-2}$.
\end{abstract}
%\date{\today}
\maketitle

\section{Introduction}

Soon after the conception of  quantum mechanics, according to which
the phase space describing a quantum system is noncommutative,
possible noncommutative (NC) structure in the space-time was also
explored \cite{Snyder}. This \nc space-times were, however, not
considered extensively until the recent re-appearance through
specific string  theory setups (for a review, see \cite{Szabo} and
references therein). The natural question which arises is what are
the phenomenological implications of such a noncommutativity, if it
exists. The natural setup  for asking this question is either the
quantum mechanics or quantum field theory on these space-times. Let
us consider  the simplest noncommutative space-time, defined by:
\be\label{Moyal} [\hat x_i,\hat x_j]=i\theta_{ij}, \ \ \ i,j=1,2,3.
\ee
Here we have assumed  \ncy only in space and kept time and space
commuting and also taken $\theta_{ij}$ to be an antisymmetric
constant matrix. Time-space noncommutativity, $\theta_{0i}\neq 0$,
leads to the well-known problems with unitarity and causality
\cite{unitarity, causality}. The above commutation relation between
the space coordinate operators will have physical effects which
could be observable in the physical experiments. One of the most
important effects is the Lorentz invariance and even the rotational
invariance violations. The noncommutative quantum field theory can
still be constructed starting from the representations of the
Poincar\'e algebra, because it has twisted-Poincar\'e symmetry, and
consequently the same representation content as a theory with usual
Poincar\'e invariance \cite{twist}. However, the violation of usual
Lorentz invariance is manifest. For example, it can change the
spectrum of hydrogen atom and have an impact on the Lamb-shift
\cite{Lamb-shift} or affect the differences between to atomic
hyperfine or Zeeman transition frequencies, in clock-comparison
experiments \cite{Kostelecky}. Using the present experimental data
one may then extract some bounds on the noncommutativity parameter
$\theta_{ij}$. It is convenient to introduce the noncommutativity
vector $\theta_i$ as
\be\label{theta} \theta_{i}=\frac{1}{2}\epsilon_{ijk}\theta_{jk} \ee
and the norm of the vector $\theta$ as inverse of the square of the
noncommutativity scale $\Lambda_{NC}$:
\be\label{NC-scale} \Lambda_{NC}^2=\frac{1}{|\vec{\theta}|}\ . \ee
The experimental {\it lower} bounds are usually represented on the
$\Lambda_{NC}$ and they are generically of order of $100-10^4$ GeV,
depending on the experiments and their precision.

The \ncy of space in the quantum field theories formulated on  \nc
space (the NCQFTs) appears through the modification of the product
of the fields which appear in the action. For the \ncy of the form
of \eqref{Moyal}, this modified product is the so-called Moyal star
product which is defined as
\be\label{star-product}
\begin{split}
(f\star g)(x) &= \exp(\frac{i}{2}\theta_{ij}\frac{\partial}{\partial x_i}\frac{\partial}{\partial y_j}) f(x)g(y)|_{x=y}\\
&= f\cdot g+\frac{i}{2}\theta_{ij}\frac{\partial f}{\partial x_i}\ \frac{\partial g}{\partial x_j}
%-\frac{1}{8}\theta_{ij}\theta_{kl} {\partial_i\partial_k}f{\partial_j\partial_l}g
+{\cal O}(\theta^2)\ ,
\end{split}
\ee
where $x$ denotes the coordinate on the commutative (Moyal)
counterpart of the noncommutative space.

In a similar manner one can construct the Yang-Mills gauge theories
on a \nc space. In particular one may construct NC $U(1)$ gauge
theory, which upon addition of fermions (electron) becomes the NCQED
\cite{NCQED-1, NCQED-2}; or one may try to construct noncommutative
versions of the electro-weak Standard Model (NCSM) \cite{NCSM,Wess}.

In the present work we would like to study one of the consequences
of \ncy when we have a sizable {\it background} electromagnetic
field.  Perhaps the most extensively studied consequence of the
presence of a constant background electric field is the Schwinger
pair production, whose \nc version has been discussed in
\cite{Chair} where it is shown that in a \nc space the pair
production in QED and NCQED are the same. The effects of a constant
background magnetic field (at tree level) does not lead to a pair production effect.
%\footnote{To be precise the background magnetic field can lead to pair production when we %consider the QED loop effects, i.e. the anomalous magnetic dipole moment  \cite{Kruglov}.}

Although the presence of a background constant magnetic field does
not have an observable effect within the tree level QED setup (or
even the Standard Model), it can have observable effects and
implications for beyond the Standard Models or at one or higher loop
levels in QED.  One of the areas where the presence of the constant
background magnetic field can be felt is the propagation of light
(photons) in models which contain an axion field \cite{P-Q}, where
we have axion production by photons propagating in a static magnetic
field (the Primakoff effect) \cite{Cameron}. The background magnetic
field can also affect propagation of photons through a photon
splitting process (${\cal F}_{\mu\nu}^4$ terms in the one-loop
effective QED) \cite{Adler}.

There have been many experiments and proposals to test the effects
of the axion, and in general the background magnetic field, most of
them focusing on the solar axions. The most famous of such
experiments is the CAST collaboration at CERN \cite{CAST}. The PLVAS
experiment \cite{PVLAS-1, PVLAS-2}, however, is an experiment set up
to test the effects of the background magnetic field on the
polarization of a linearly polarized photon, effects which may be
caused by a terrestrial axion.

The PVLAS collaboration has been able to observe a rotation of the
polarization of light while traveling through a magnetic field {\it
transverse} to the direction of photon propagation. For such a
background magnetic field the photon splitting effects are basically
absent. Moreover, as has been reported \cite{PVLAS-1, PVLAS-2},
using the bounds on the  axion parameters coming from the solar
experiments it is not possible to explain the observed polarization
rotation. (The bounds would only allow for polarization rotation
smaller than the observed ones by seven orders of magnitude.)

In this work we provide an alternative explanation for the PVLAS
results. Namely, we will show that the \ncy effects in the
background magnetic field  can cause a change in the dispersion
relation and hence the polarization rotation. These \nc effects,
however, are not going to change the CAST experiment or similar
experiments which are built upon "Light Shining Through a Wall"
\cite{LSTW} computations.

The rest of the paper is organized as follows. First we present the
action of the \nc $U(1)$ Yang-Mills gauge theory and work out the
modified energy-momentum dispersion relation in this setup. We then
compute the polarization rotation in an external magnetic field and
show that the PVLAS experiment results can be explained within this
\nc model with the \ncy scale $\Lambda_{NC}$ around $30 $ GeV.

%\section{Noncommutative $U(1)$ Yang-Mills Theory}

\section{Propagation of a photon in a \nc space in background magnetic field}

We start with the action of a NC $U(1)$ gauge theory:
\be\label{action-0}
S=-\frac{1}{4\pi}\int d^4 x {\cal F}^{\mu\nu}\star {\cal F}_{\mu\nu}
\ee
where
\be\label{calF} {\cal F}_{\mu\nu}=\partial_\mu {\cal
A}_\nu-\partial_\mu {\cal A}_\nu+ie[{\cal A}_\mu,{\cal
A}_\nu]_{\star} \ee and $[{\cal A}_\mu,{\cal A_\nu}]_\star$ is the
Moyal bracket, defined as
\[
[{\cal A}_\mu,{\cal A_\nu}]_\star={\cal A}_\mu \star{\cal A_\nu} - {\cal A}_\nu \star{\cal A_\mu}.
\]

In our case, when there is a background magnetic field turned on,
\be\label{calA} {\cal A}_\mu=A^0_{\mu}+A_\mu \ee
where $A^0_\mu$ is the background gauge field and $A_\mu$ is the
gauge field corresponding to the propagating light.

Plugging \eqref{calA} into \eqref{calF}, we have
\[
{\cal F}_{\mu\nu}=F^0_{\mu\nu}+ F_{\mu\nu}
\]
where $F^0_{\mu\nu}$ is the background field strength:
\be\label{backgroundF} F^0_{\mu\nu}= \partial_\mu
A^0_\nu-\partial_\mu A^0_\nu+ie[ A^0_\mu,A^0_\nu]_{\star}\ .\ee
Since in the PVLAS experiment the applied background is a constant
magnetic field, $\vec B_0$, it follows that
$$
A^0_0=0
$$
and $A^0_i$ is found from the relation of definition of $\vec B_0$,
i.e.
\be F^0_{ij}=\epsilon_{ijk} B^k_0, \ee
leading to the equation
\be\label{backgroundB}
\partial_i
A^0_j-\partial_j A^0_i+ie[ A^0_i,A^0_j]_{\star}= \epsilon_{ijk}
B^k_0\ .\ee
Solving (\ref{backgroundB}) for $A^0_i$, we obtain the gauge
potential of the applied background field as a power series in
$\theta$ (in effect, a power series in $ e\
\vec\theta\cdot\vec B_0$):
\be\label{backgroundA} A_i^0=\frac{1}{2}\epsilon_{ijk} B^j_0\
S\left(e\ \vec\theta\cdot\vec B_0\right) x^k \ee
where the power series $S\left(e\ \vec\theta\cdot\vec B_0\right)$ is
obtained as solution of the equation
$$
\left(\frac{e}{4} \vec \theta\cdot\vec B_0\right)\ S^2 - S + 1 = 0.
$$
From (\ref{backgroundA}) we can extract an effective background
magnetic field $\vec B$, which is related to the applied magnetic
field $\vec B_0$ as
\be\label{BvsB_0} \vec B=\vec B_0\ S\left(e\ \vec\theta\cdot\vec
B_0\right)\ .\ee

In the $A_0=0$ gauge, we have:
\[\begin{split}
{\cal F}_{0i}&=\partial_0A_i\ ,\cr
{\cal F}_{ij}&=\partial_{[i}A_{j]}+ie\left([A_i^0,A_j]_\star+[A_i,A_j^0]_\star\right)+{\cal O}(A^2)\\
=&(1-\frac{e}{2}{\vec{\theta}}\cdot{\vec{B}})\partial_{[i}A_{j}+\frac{e}{2}
(\theta_i \vec{B}\cdot {\vec{\partial}}A_j- i\leftrightarrow j)
%\theta_j \vec{B}_0\cdot {\vec{\partial}}A_i)
+{\cal O}(A^2)
\end{split}\]

Inserting the above expressions for the components of ${\cal
F}_{\mu\nu}$ into the action \eqref{action-0}, and dropping the
parts only involving the background $B_0$ field and the total
derivative terms we obtain
\[
S=\frac{1}{4\pi}\int d^4x\left(2(\partial_0 A_i)^2- {\cal
F}_{ij}^2\right)
\]
and hence the equations of motion are
\be\label{EOM-A}
\begin{split}
\biggl[&\delta_{ij}\left(\partial_t^2-(1-\frac{1}{2}e{\vec{\theta}}\cdot{\vec{B}})^2
\nabla^2\right)\\
&-e(1-\frac{1}{2}e{\vec{\theta}}\cdot{\vec{B}})(\delta_{ij}({\vec{\theta}}\cdot\nabla)-\frac{1}{2}\theta_j
\partial_i){\vec{B}}\cdot\nabla\\
&-\frac{e^2}{4}(\theta^2\delta_{ij}-\theta_i\theta_j)
(\vec{B}\cdot\nabla)^2\biggr]A_j(x)=0
\end{split}
\ee
Expanding
\be\label{photon-wave}
A_i(x)=\int d^3k\ \epsilon_{i}(k) e^{i(\omega t-\vec{k}\cdot\vec{x})}
\ee
we obtain
\be\label{EOM-eps}
\begin{split}
\biggl[&\delta_{ij}\left(\omega^2-(1-\frac{1}{2}e{\vec{\theta}}\cdot{\vec{B}})^2
k^2\right)\\
&-e(1-\frac{1}{2}e{\vec{\theta}}\cdot{\vec{B}}){\vec{B}}\cdot{\vec{k}}({\vec{\theta}}\cdot{\vec{k}}\delta_{ij}-
\frac{1}{2}\theta_j k_i)\\
&-\frac{e^2}{4}(\theta^2\delta_{ij}-\theta_i\theta_j)
(\vec{B}\cdot{\vec{k}})^2\biggr]\epsilon_j(k)=0\ .
\end{split}
\ee
Next we note that in the PVLAS setup the external magnetic field is
transverse to the direction of the light beam propagation, i.e.
$\vec{B}_0\cdot \vec{k}=0$. For this case, after imposing the
transversality condition which is necessary to fix the remainder of
the NC $U(1)$ gauge freedom,  the equation of motion simplifies
considerably and leads to the following modified dispersion relation
\be\label{dis-rel}
\omega^2=(1-\frac{1}{2}e{\vec{\theta}}\cdot{\vec{B}})^2 k^2 \ee

To relate the above dispersion relation to the rotation of the polarization vector
%rotation reported in PVLAS experiment
we note that the rotation of the polarization resulting from the
change in the dispersion relation of the form
\[
\omega-k= k\Delta
\]
is
\be\label{pol-rot} \delta \phi= kL\Delta\ , \ee
where $L$ is the length the photon has passed through the external
magnetic field (or the length of the Fabry-Perot cavity in the PVLAS
setup). In our model
\be\label{delta} \Delta=\frac{1}{2}e{\vec{\theta}}\cdot{\vec{B}} \ee

Putting \eqref{pol-rot} and \eqref{delta} together we obtain
\be\label{Pol-Rot} \delta\phi=\pi \frac{L}{\lambda}\
e{\vec{\theta}}\cdot{\vec{B}}. \ee
Recall that $\vec B$ is not exactly the applied background magnetic
field. The effective background magnetic field $\vec B$, in our
notations, is given in \eqref{BvsB_0} and slightly differs from
$B_0$, by a power series in $\left(e\ \vec\theta\cdot\vec
B_0\right)$. For $\left(e\ \vec\theta\cdot\vec B_0\right)$ small,
however, we can discard the higher-order terms in the power series
and practically treat the background field as a commutative object.

\section{Comparison to the PVLAS results}

Now that we have shown and computed the polarization rotation in the
background magnetic field in the \nc setup we are ready to compare
our result with the PVLAS experiment. In the PVLAS experiment the
background $B_0$ field is on a turntable which rotates with a
frequency of $\nu_{B}=0.33\ {\rm Hz}$ and the magnitude of $B_0=5.5\
{\rm T}$. In our computations we have ignored the time dependence of
the external magnetic field and that the magnet is placed on a
turn-table. This will, however, not alter our result, if used for
the PVLAS experiment, as the frequency of the light beam $\omega\sim
10^5\ {\rm GHz}$ is much larger than that of the magnetic field
$\nu_B\sim 0.1\ {\rm Hz}$ and hence in our revolution of the photon
wave the magnetic field is essentially a constant.
%Now we may consider the time dependence of the background magnetic
%field.
Moreover, as is seen from \eqref{delta} the rotation in the
polarization is in phase with  the background magnetic field.

In the PVLAS experiment \cite{PVLAS-2}:
\be\label{PVLAS-data}
\begin{split}
B_0&= 5.5\ {\rm T} , \ \ \delta\phi=(2.2\pm 0.3)\times 10^{-7}\ {\rm
rad}\cr \lambda &=1064\ {\rm nm}\ , \ \ \ \ \  L= N\cdot l_0\ ,
\end{split}
\ee where $N$ is the number of passes through the magnetic field
region which has length $l_0$. In the PVLAS experiment,
$N=\frac{2{\cal F}}{\pi}$ with the finesse parameter ${\cal
F}=8.2\times 10^{5}$, leading to $N=5.22\times 10^{5}$, and
$l_0=1.333\ {\rm m}$.

Noting that $eB_0= 3.25\times 10^{-10} {\rm MeV}^2$ for a magnetic
field of 5.5 T, defining the noncommutativity scale $\Lambda_{NC}$
as in (\ref{NC-scale}) and assuming that $\vec{\theta}$ is parallel
to $\vec{B}_0$ we obtain
\be\label{NC-scale-PVLAS} \Lambda_{NC}\simeq 30\ {\rm GeV}. \ee
\eqref{NC-scale-PVLAS} is our main result.

\section{Discussions and conclusions}

The value obtained for the \ncy energy scale from the PVLAS data,
\eqref{NC-scale-PVLAS}, is by two orders of magnitude lower than the
one obtained from other considerations, such as NC Lamb shift
\cite{Lamb-shift}, clock comparison experiments \cite{Kostelecky} or
precision data of Standard Model \cite{NCSM}. As such, \ncy cannot
explain the amount of the polarization rotation in the PVLAS
experiment. Of course, it is also possible that the \ncy is not the
only source for the polarization rotation measured in the PVLAS
experiment.

However, the interpretation of the PVLAS data as due to \ncy is
favoured, as compared to the interpretation in terms of axion
effects. As mentioned in the introduction, assuming that the solar
and terrestrial axion parameters, as the most natural case, are the
same, the PVLAS results cannot be caused by the axions, since there
are seven orders of magnitude difference between the PVLAS results
and the estimation of the polarization rotation from the CAST data.
As the PVLAS data look at present, the amount of the polarization
rotation cannot be explained by any known physics beyond the
Standard Model.

An important point is that the bounds on the \ncy scale discussed in
the literature are all based so far on the ``non-observation'' of
the \nc effects. In this sense, PVLAS might serve as the first
experiment in which the \nc effects are observed. Thus space-time
noncommutativity could be a plausible candidate, would the improved
PVLAS data change to sufficiently lower values. As such, an
improvement in the PVLAS results, which we are awaiting, would shed
light on the \ncy of the space-time.

\centerline{{\bf Acknowledgments}} We would like to thank Per Osland
for asking the right question. The financial support of the Academy
of Finland under the Projects No. 54023 and 104368 is greatly
acknowledged.

%\end{acknowledgments}


\begin{thebibliography}{99}

%%
%%  bibliographic items can be constructed using the LaTeX format in SPIRES:
%%    see    http://www.slac.stanford.edu/spires/hep/latex.html
%%  SPIRES will also supply the CITATION line information; please include it.
%%

%\bibitem{}

%\bibitem{}

\bibitem{Snyder}
  H.~S.~Snyder,
  ``Quantized Space-Time,''
  Phys.\ Rev.\  {\bf 71} (1947) 38.
  %%CITATION = PHRVA,71,38;%%

 C.~N.~Yang,
  ``On Quantized Space-Time,''
  Phys.\ Rev.\  {\bf 72} (1947) 874.
  %%CITATION = PHRVA,72,874;%%

\bibitem{Szabo}
  R.~J.~Szabo,
  ``Quantum field theory on noncommutative spaces,''
  Phys.\ Rept.\  {\bf 378}, 207 (2003)
  [arXiv:hep-th/0109162].
  %%CITATION = HEP-TH 0109162;%%

\bibitem{unitarity}
J.~Gomis and T.~Mehen, ``Space-time noncommutative field theories
and unitarity,'' Nucl.\ Phys.\ B {\bf 591}, 265 (2000)
[arXiv:hep-th/0005129].
%%CITATION = HEP-TH 0005129;%%

\bibitem{causality}
N.~Seiberg, L.~Susskind and N.~Toumbas, ``Space/time
non-commutativity and causality,'' JHEP {\bf 0006}, 044 (2000)
[arXiv:hep-th/0005015].
%%CITATION = HEP-TH 0005015;%%

M.~Chaichian, K.~Nishijima and A.~Tureanu, ``Spin-statistics and CPT
theorems in noncommutative field theory,'' Phys.\ Lett.\ B {\bf
568}, 146 (2003) [arXiv:hep-th/0209008].
%%CITATION = HEP-TH 0209008;%%




\bibitem{twist}
M.~Chaichian, P.~P.~Kulish, K.~Nishijima and A.~Tureanu, ``On a
Lorentz-invariant interpretation of noncommutative space-time and
its implications on noncommutative QFT,'' Phys.\ Lett.\ B {\bf 604},
98 (2004) [arXiv:hep-th/0408069].
%%CITATION = HEP-TH 0408069;%%



\bibitem{Lamb-shift}

  M.~Chaichian, M.~M.~Sheikh-Jabbari and A.~Tureanu,
  ``Hydrogen atom spectrum and the Lamb shift in noncommutative QED,''
  Phys.\ Rev.\ Lett.\  {\bf 86}, 2716 (2001)
  [arXiv:hep-th/0010175].
  %%CITATION = HEP-TH 0010175;%%


\bibitem{Kostelecky}
S.~M.~Carroll, J.~A.~Harvey, V.~A.~Kostelecky, C.~D.~Lane and
T.~Okamoto, ``Noncommutative field theory and Lorentz violation,''
Phys.\ Rev.\ Lett.\  {\bf 87}, 141601 (2001) [arXiv:hep-th/0105082].
%%CITATION = HEP-TH 0105082;%%



\bibitem{NCQED-1}
  M.~Hayakawa,
 ``Perturbative analysis on infrared aspects of noncommutative QED on  $R^4$,''
  Phys.\ Lett.\ B {\bf 478}, 394 (2000)
  [arXiv:hep-th/9912094];
  %%CITATION = HEP-TH 9912094;%%
  % M.~Hayakawa,
 ``Perturbative analysis on infrared and ultraviolet aspects of
  noncommutative QED on $R^4$,''
  [arXiv:hep-th/9912167].
  %%CITATION = HEP-TH 9912167;%%

 \bibitem{NCQED-2}

%  M.~M.~Sheikh-Jabbari,
%  ``Discrete symmetries (C,P,T) in noncommutative field theories,''
%  Phys.\ Rev.\ Lett.\  {\bf 84}, 5265 (2000)
%  [arXiv:hep-th/0001167].
  %%CITATION = HEP-TH 0001167;%%

  I.~F.~Riad and M.~M.~Sheikh-Jabbari,
  ``Noncommutative QED and anomalous dipole moments,''
  JHEP {\bf 0008}, 045 (2000)
  [arXiv:hep-th/0008132].
  %%CITATION = HEP-TH 0008132;%%




\bibitem{NCSM}
 M.~Chaichian, P.~Pre\v{s}najder, M.~M.~Sheikh-Jabbari and A.~Tureanu,
  ``Noncommutative standard model: Model building,''
  Eur.\ Phys.\ J.\ C {\bf 29}, 413 (2003)
  [arXiv:hep-th/0107055].
  %%CITATION = HEP-TH 0107055;%%

\bibitem{Wess}
X.~Calmet, B.~Jur\v{c}o, P.~Schupp, J.~Wess and M.~Wohlgenannt,
``The standard model on non-commutative space-time,'' Eur.\ Phys.\
J.\ C {\bf 23}, 363 (2002) [arXiv:hep-ph/0111115].
%%CITATION = HEP-PH 0111115;%%





\bibitem{Chair}
  L.~\'Alvarez-Gaum\'e and J.~L.~F.~Barbon,
 ``Non-linear vacuum phenomena in non-commutative QED,''
  Int.\ J.\ Mod.\ Phys.\ A {\bf 16}, 1123 (2001)
  [arXiv:hep-th/0006209].
  %%CITATION = HEP-TH 0006209;%%

  N.~Chair and M.~M.~Sheikh-Jabbari,
  ``Pair production by a constant external field in noncommutative QED,''
  Phys.\ Lett.\ B {\bf 504}, 141 (2001)
  [arXiv:hep-th/0009037].
  %%CITATION = HEP-TH 0009037;%%

%\bibitem{Kruglov}
%  S.~I.~Kruglov,
%  ``Pair production and solutions of the wave equation for particles with
%  arbitrary spin,''
%  arXiv:hep-ph/9908410.
  %%CITATION = HEP-PH 9908410;%%

\bibitem{P-Q}
  R.~D.~Peccei and H.~R.~Quinn,
  ``CP conservation in the presence of instantons,''
  Phys.\ Rev.\ Lett.\  {\bf 38}, 1440 (1977).
  %%CITATION = PRLTA,38,1440;%%

\bibitem{Cameron}
  R.~Cameron {\it et al.},
  ``Search for nearly massless, weakly coupled particles by optical
  techniques,''
  Phys.\ Rev.\ D {\bf 47} (1993) 3707.
  %%CITATION = PHRVA,D47,3707;%%

\bibitem{Adler}
S.~L.~Adler,
  ``Photon splitting and photon dispersion in a strong magnetic field,''
  Annals Phys.\  {\bf 67}, 599 (1971).
  %%CITATION = APNYA,67,599;%%

\bibitem{CAST}
  S.~Andriamonje {\it et. al.}, %  [CAST Collaboration],
  ``First results from the CERN axion solar telescope (CAST),''
  Phys. Rev. Lett.  {\bf 94}, 121301 (2005)
  [arXiv:hep-ex/0411033].
  %%CITATION = HEP-EX 0411033;%%
For more information about the CERN Axion Solar Telescope (CAST) collaboration visit the
 website: http://cast.web.cern.ch/CAST/



\bibitem{PVLAS-1}
  E.~Zavattini {\it et al.}  [PVLAS Collaboration],
  ``Experimental observation of optical rotation generated in vacuum by a
  magnetic field,''
  [arXiv:hep-ex/0507107].
  %%CITATION = HEP-EX 0507107;%%


\bibitem{PVLAS-2}
 U.~Gastaldi,
 ``PVLAS results'',
 Presented at 40th Rencontres de Moriond on Electroweak Interactions and Unified Theories,
 % La Thuile, Aosta Valley, Italy, 5-12
  March 2005, arXiv:hep-ex/0507061.
  %%CITATION = HEP-EX 0507061;%%

\bibitem{LSTW}
  K.~Van Bibber, N.~R.~Dagdeviren, S.~E.~Koonin, A.~Kerman and H.~N.~Nelson,
  ``An experiment to produce and detect light pseudoscalars,''
  Phys.\ Rev.\ Lett.\  {\bf 59}, 759 (1987).
  %%CITATION = PRLTA,59,759;%%







\end{thebibliography}
\end{document}